\documentclass[twocolumn,twoside]{article}
\usepackage{graphics}
\usepackage[english]{babel}

\newcommand{\hb}{\\ \hspace*{2ex}}

\begin{document}
\title{2-field model of dark energy \\with canonical and non-canonical kinetic terms}
\author{O.\,Sergijenko\\[2mm] 
 Astronomical Observatory, Taras Shevchenko National University of Kyiv,\hb
 Observatorna str. 3, Kyiv, 04053, Ukraine,  {\em olga.sergijenko.astro@gmail.com}\\[2mm]
}
\date{}
\maketitle

ABSTRACT. We generalize quintom to include the tachyonic kinetic term along with the classical one. For such a model we obtain the expressions for energy density and pressure. For the spatially flat, homogeneous and isotropic Universe with Friedmann-Robertson-Walker metric of 4-space we derive the equations of motion for the fields. We discuss in detail the reconstruction of the scalar fields potential $U(\phi,\xi)$. Such a reconstruction cannot be done unambiguously, so we consider 3 simplest forms of $U(\phi,\xi)$: the product of $\Phi(\phi)$ and $\Xi(\xi)$, the sum of $\Phi(\phi)$ and $\Xi(\xi)$ and this sum to the $\kappa$th power.\\[1mm]
\\[2mm]

{\bf 1. Introduction}\\[1mm]

20 years ago the accelerated expansion of the Universe was discovered. The cosmological constant in Einstein equations (equation of state parameter $w=-1$) is the simplest explanation of it. The dark energy in form of a scalar field is the most popular alternative to $\Lambda$. Its equation of state parameter can either be constant or vary in time. Dark energy with $w>-1$ is called quintessence, with $w<-1$ -- phantom. For several widely used parametrizations of $w(z)$, e.g. Chevallier \& Polarski (2001) and Linder (2003) (CPL), Komatsu (2009) (WMAP5), at a certain redshift there is transition from quintessence to phantom or vice versa. Such behavior is forbidden for a single minimally coupled scalar field, as it was shown for the first time by Vikman A. (2005) (see also Easson D.A. \& Vikman A. (2016)). However, crossing of the phantom divide is possible in the cases of kinetic gravity braiding (Deffayet C., Pujolas O., Sawicki I., Vikman A. (2010)), sound speed vanishing in phantom domain (Creminelli P., D'Amico G., Norena J., Vernizzi F. (2009)) or non-minimal couplings (Amendola L. (2000), Pettorino V. \& Baccigalupi C. (2008)).

The most popular scalar field model allowing for $w=-1$ crossing is quintom proposed by Feng B., Wang X., Zhang X. (2005). It is the 2-field model with 2 canonical kinetic terms -- one with the ``+'' sign for quintessence and one with the ``-'' sign for phantom -- and a potential $U(\phi,\xi)$ in Lagrangian.

Quintom can be generalized to include a non-canonical kinetic term. The simplest physically motivated Lagrangian with the non-canonical kinetic term is the tachyon one. So, we propose the 2-field model of dark energy with classical and tachyonic kinetic terms.
\\[2mm]

{\bf 2. 2-field model with classical and tachyonic kinetic terms}\\[1mm]

We consider the spatially flat, homogeneous and isotropic Universe with Friedmann-Robertson-Walker (FRW) metric of 4-space
\begin{eqnarray}
ds^2=g_{ij} dx^i dx^j =a^2(\eta)(d\eta^2-\delta_{\alpha\beta} dx^{\alpha}dx^{\beta})\label{metr}
\end{eqnarray}
(here $i,j=0,1,2,3$, $\alpha,\beta=1,2,3$, $a$ is the scale factor, $\eta$ is the conformal time and $c=1$). The Universe is filled with relativistic (radiation, neutrinos), non-relativistic (baryons, dark matter) matter and dark energy. The latter is modeled by 2 scalar fields with the Lagrangian:
\begin{eqnarray}
L=-X_{\phi}-U(\phi,\xi)\sqrt{1-2X_{\xi}},\label{lagr}
\end{eqnarray}
where
\begin{eqnarray*}
&&X_{\phi}=\frac{1}{2}\phi_{,i}\phi^{,i}=\frac{\dot{\phi}^2}{2},\\
&&X_{\xi}=\frac{1}{2}\xi_{,i}\xi^{,i}=\frac{\dot{\xi}^2}{2}
\end{eqnarray*}
are the kinetic terms. This Lagrangian is classical (Klein-Gordon) with respect to the field $\phi$ which corresponds to phantom and tachyon (Dirac-Born-Infeld) with respect to the field $\xi$ which corresponds to quintessence.

The energy density and pressure for such a model are as follows:
\begin{eqnarray}
&&\rho_{de}=-X_{\phi}+\frac{U(\phi,\xi)}{\sqrt{1-2X_{\xi}}},\label{rho}\\
&&p_{de}=-X_{\phi}-U(\phi,\xi)\sqrt{1-2X_{\xi}}.\label{p}
\end{eqnarray}
The dark energy equation of state (EoS) parameter is defined as $w(a)=p_{de}/\rho_{de}$.

For the metric (\ref{metr}) the Lagrangian (\ref{lagr}) yields the following equations of motion:
\begin{eqnarray}
&&\ddot{\phi}+2aH\dot{\phi}-\frac{\partial U}{\partial\phi}a^2\sqrt{1-\frac{\dot{\xi}^2}{a^2}}=0,\\
&&\ddot{\xi}+2aH\dot{\xi}-3aH\dot{\xi}\frac{\dot{\xi}^2}{a^2}\nonumber\\
&&+\frac{1}{U}\left(\frac{\partial U}{\partial\phi}\dot{\phi}\dot{\xi}+a^2\frac{\partial U}{\partial\xi}\right)\left(1-\frac{\dot{\xi}^2}{a^2}\right)=0.
\end{eqnarray}
Here a dot denotes the derivative with respect to $\eta$ and $H\equiv\dot{a}/a$ is the Hubble parameter.

From (\ref{rho})-(\ref{p}) it is clear that reconstruction of the potential $U(\phi,\xi)$ cannot be done unambiguously and requires additional assumptions. First of all, we should choose a form of $U(\phi,\xi)$. We restrict our consideration to 3 simplest ansatzes from Andrianov A.A., Cannata F., Kamenshchik A.Y., Regoli D. (2008): 
\begin{itemize}
\item $U(\phi,\xi)=\Phi(\phi)\Xi(\xi)$,
\item $U(\phi,\xi)=\Phi(\phi)+\Xi(\xi)$ and
\item $U(\phi,\xi)=[\Phi(\phi)+\Xi(\xi)]^{\kappa},\;\kappa=const$.
\end{itemize}

Secondly, we should assume either $X_{\phi}$ or $X_{\xi}$ to be a known function of the scale factor. Then for $X_{\xi}=\alpha(a)$ we get:
\begin{eqnarray}
&&U(a)=\frac{1}{2}\frac{1-w}{1-\alpha}\sqrt{1-2\alpha}\rho,\label{uaa}\\
&&X_{\phi}(a)=-\frac{1}{2}\frac{1-2\alpha+w}{1-\alpha}\rho\label{xaa}
\end{eqnarray}
and for $X_{\phi}=\beta(a)$:
\begin{eqnarray}
&&U(a)=\sqrt{-(\rho w+\beta)(\rho+\beta)},\label{uab}\\
&&X_{\xi}(a)=\frac{1}{2}\frac{\rho(1+w)+2\beta}{\rho+\beta}.\label{xab}
\end{eqnarray}
Dependences of the fields $\phi$ and $\xi$ on the scale factor $a$ are determined from $X_{\xi}=\alpha$ and (\ref{xaa}):
\begin{eqnarray}
&&\xi(a)=\pm\int\frac{da}{aH}\sqrt{2\alpha},\label{xia}\\
&&\phi(a)=\pm\int\frac{da}{aH}\sqrt{-\frac{1}{2}\frac{1-2\alpha+w}{1-\alpha}\rho}\label{phia}
\end{eqnarray}
or $X_{\phi}=\beta$ and (\ref{xab}):
\begin{eqnarray}
&&\phi(a)=\pm\int\frac{da}{aH}\sqrt{2\beta},\label{phib}\\
&&\xi(a)=\pm\int\frac{da}{aH}\sqrt{\frac{1}{2}\frac{\rho(1+w)+2\beta}{\rho+\beta}}.\label{xib}
\end{eqnarray}
These expressions together with either (\ref{uaa}) or (\ref{uab}) define $U(\phi)$ in the parametric form. This allows us to reconstruct the potential even if the integrals in (\ref{xia})-(\ref{xib}) cannot be solved analytically.\\[2mm]

{\bf 3. Reconstructed potentials in an explicit form}\\[1mm]

If it is possible to invert the analytical dependences (\ref{xia})-(\ref{phia}) or (\ref{phib})-(\ref{xib}) then we can obtain the explicit expressions for potentials (as it has been done for single-field models in e.g. Sergijenko \& Novosyadlyj (2008), Novosyadlyj \& Sergijenko (2009)). For $U(\phi,\xi)=\Phi(\phi)\Xi(\xi),\;X_{\xi}=\alpha$ the potential is reconstructed as:
\begin{eqnarray*}
&&U(\phi,\xi)=\exp\left\{\pm\int d\phi\left[\frac{1}{a}\sqrt{-\frac{(1-\alpha)(1-2\alpha+w)}{\rho}}\right.\right.\\
&&\left.\left.\times\frac{1}{(1-w)(1-1-2\alpha)}\left(\frac{\dot{w}-2\dot{\alpha}}{1-2\alpha+w}+\frac{\dot{\alpha}}{1-\alpha}\right.\right.\right.\\
&&\left.\left.\left.+3aH(1-w)\right)\right](a(\phi))\mp\int d\xi\left[\frac{1}{a}\frac{\sqrt{2\alpha}}{1-2\alpha}\left(\frac{1}{2}\frac{\dot{\alpha}}{\alpha}\right.\right.\right.\\
&&\left.\left.\left.+\frac{\dot{\alpha}}{1-\alpha}\frac{1-2\alpha+w}{1-w}+\frac{\dot{w}-2\dot{\alpha}}{1-w}\right.\right.\right.\\
&&\left.\left.\left.+3aH(2-4\alpha+w)\right)\right](a(\xi))\right\},
\end{eqnarray*}
for $U(\phi,\xi)=\Phi(\phi)\Xi(\xi),\;X_{\phi}=\beta$ it is as follows:
\begin{eqnarray*}
&&U(\phi,\xi)=\exp\left\{\pm\int d\phi\left[\frac{1}{a}\frac{\sqrt{2\beta}}{\rho w+\beta}\left(3aH+\frac{1}{2}\frac{\dot{\beta}}{\beta}\right)\right]\right.\\
&&\left.\times(a(\phi))\pm\int d\xi\left[\frac{1}{2a}\frac{\sqrt{(\rho(1+w)+2\beta)(\rho+\beta)}}{\rho w+\beta}\right.\right.\\
&&\left.\left.\times\left(\frac{\rho(\dot{w}-3aH(1+w)^2)+2\dot{\beta}}{\rho(1+w)+2\beta}\right.\right.\right.\\
&&\left.\left.\left.+3\frac{aH(1-w)\rho-6aH\beta-\dot{\beta}}{\rho+\beta}\right)\right](a(\xi))\right\}.
\end{eqnarray*}
For $U(\phi,\xi)=\Phi(\phi)+\Xi(\xi),\;X_{\xi}=\alpha$ we get:
\begin{eqnarray*}
&&U(\phi,\xi)=\pm\int d\phi\left[\frac{1}{2a}\sqrt{-\frac{1-2\alpha+w}{(1-\alpha)(1-2\alpha)}\rho}\right.\\
&&\left.\times\left(\frac{\dot{w}-2\dot{\alpha}}{1-2\alpha+w}+\frac{\dot{\alpha}}{1-\alpha}+3aH(1-w)\right)\right](a(\phi))\\
&&\mp\int d\xi\left[\frac{1}{2a}\sqrt{\frac{2\alpha}{1-2\alpha}}\frac{1-w}{1-\alpha}\rho\left(\frac{1}{2}\frac{\dot{\alpha}}{\alpha}+\frac{\dot{\alpha}}{1-\alpha}\right.\right.\\
&&\left.\left.\times\frac{1-2\alpha+w}{1-w}+\frac{\dot{w}-2\dot{\alpha}}{1-w}+3aH\right.\right.\\
&&\left.\left.\times(2-4\alpha+w)\right)\right](a(\xi)),
\end{eqnarray*}
while for $U(\phi,\xi)=\Phi(\phi)+\Xi(\xi),\;X_{\phi}=\beta$:
\begin{eqnarray*}
&&U(\phi,\xi)=\pm\int d\phi\left[\frac{1}{a}\sqrt{2\beta}\sqrt{-\frac{\rho+\beta}{\rho w+\beta}}\left(3aH\right.\right.\\
&&\left.\left.+\frac{1}{2}\frac{\dot{\beta}}{\beta}\right)\right](a(\phi))\pm\int d\xi\left[\frac{1}{2a}\sqrt{-\frac{\rho(1+w)+2\beta}{\rho w+\beta}}\right.\\
&&\left.\times(\rho+\beta)\left(\frac{\rho(\dot{w}-3aH(1+w)^2)+2\dot{\beta}}{\rho(1+w)+2\beta}\right.\right.\\
&&\left.\left.+3\frac{aH(1-w)\rho-6aH\beta-\dot{\beta}}{\rho+\beta}\right)\right](a(\xi)).
\end{eqnarray*}

In the case of $U(\phi,\xi)=[\Phi(\phi)+\Xi(\xi)]^{\kappa},\;\kappa=const,\;X_{\xi}=\alpha$ the potential reads:
\begin{eqnarray*}
&&U(\phi,\xi)=\frac{1}{(2\kappa)^{\kappa}}\left\{\pm\int d\phi\left[\frac{1}{a}\sqrt{-(1-2\alpha+w)}\rho^{\frac{1}{\kappa}-\frac{1}{2}}\right.\right.\\
&&\left.\left.\times(1-w)^{\frac{1-\kappa}{\kappa}}(1-\alpha)^{\frac{1}{2}-\frac{1}{\kappa}}(1-2\alpha)^{\frac{1}{2\kappa}-1}\left(\frac{\dot{w}-2\dot{\alpha}}{1-2\alpha+w}\right.\right.\right.\\
&&\left.\left.\left.+\frac{\dot{\alpha}}{1-\alpha}+3aH(1-w)\right)\right](a(\phi))\mp\int d\xi\left[\frac{1}{a}\sqrt{2\alpha}\right.\right.\\
&&\left.\left.\times(1-2\alpha)^{\frac{1}{2\kappa}-1}\left(\frac{1-w}{1-\alpha}\rho\right)^{\frac{1}{\kappa}}\left(\frac{1}{2}\frac{\dot{\alpha}}{\alpha}+\frac{\dot{\alpha}}{1-\alpha}\right.\right.\right.\\
&&\left.\left.\left.\times\frac{1-2\alpha+w}{1-w}+\frac{\dot{w}-2\dot{\alpha}}{1-w}+3aH(2-4\alpha+w)\right)\right]\right.\\
&&\left.\times(a(\xi))\right\}
\end{eqnarray*}
and $U(\phi,\xi)=[\Phi(\phi)+\Xi(\xi)]^{\kappa},\;\kappa=const,\;X_{\phi}=\beta$ yields:
\begin{eqnarray*}
&&U(\phi,\xi)=\frac{1}{\kappa^{\kappa}}\left\{\pm\int d\phi\left[\frac{1}{a}\sqrt{2\beta}(\rho+\beta)^{\frac{1}{2\kappa}}\right.\right.\\
&&\left.\left.\times(-(\rho w+\beta))^{\frac{1}{2\kappa}-1}\left(3aH+\frac{1}{2}\frac{\dot{\beta}}{\beta}\right)\right](a(\phi))\right.\\
&&\left.\mp\int d\xi\left[\frac{1}{2a}\sqrt{\rho(1+w)+2\beta}(\rho+\beta)^{\frac{\kappa+1}{2\kappa}}\right.\right.\\
&&\left.\left.\times(-(\rho w+\beta))^{\frac{1}{2\kappa}-1}\left(\frac{\rho(\dot{w}-3aH(1+w)^2)+2\dot{\beta}}{\rho(1+w)+2\beta}\right.\right.\right.\\
&&\left.\left.\left.+3\frac{aH(1-w)\rho-6aH\beta-\dot{\beta}}{\rho+\beta}\right)\right](a(\xi))\right\}.
\end{eqnarray*}

It is clear that for the multicomponent cosmological models with realistic parametrizations of the dark energy EoS crossing $-1$ (CPL, WMAP5) these expressions are not suitable for practical use since even for $\alpha=const$ or $\beta=const$ the integrals cannot be solved analytically. Thus, to reconstruct the potentials for certain values of cosmological parameters and dependences $w(a)$ and $\alpha(a)$ or $\beta(a)$  we have to use the parametric dependences (\ref{uaa}), (\ref{xia}), (\ref{phia}) or (\ref{uab}), (\ref{phib}), (\ref{xib}) which define the potential $U(\phi,\xi)$.\\[2mm]

{\bf 4. Conclusion}\\[1mm]

In the reconstruction of potential of the proposed 2-field model of dark energy with classical and tachyonic kinetic terms there are 2 ambiguities requiring additional assumptions: about the form of $U(\phi,\xi)$ and about the dependence of a kinetic term on the scale factor. The third ambiguity -- a sign in front of the integrals in (\ref{xia})-(\ref{xib}) -- is less important since the potentials with ``+'' and ``-'' are symmetric with respect to $\phi=0$ and $\xi=0$. For 3 ansatzes for the form of $U(\phi,\xi)$ we obtained the reconstructed potentials in explicit form and in parametric one that is suitable for multicomponent cosmological models and all parametrizations of the dark energy equation of state.\\[2mm]

{\it Acknowledgements.} This work has been supported in part by the Department of target training of Taras Shevchenko National University of Kyiv under National Academy of Sciences
of Ukraine (project 6$\Phi$).
\\[3mm]
{\bf References\\[2mm]}
Andrianov A.A., Cannata F., Kamenshchik A.Y., Regoli D.: 2008, {\it J. Cosmol. Astropart. Phys.}, {\bf 02}, 015.\\
Amendola L.: 2000, {\it Phys. Rev. D}, {\bf 62}, 043511.\\
Chevallier M. \& Polarski D.: 2001, {\it Int. J. Mod. Phys. D}, {\bf 10}, 213.\\
Creminelli P., D'Amico G., Norena J., Vernizzi F.: 2009, {\it J. Cosmol. Astropart. Phys.}, {\bf 02}, 018.\\
Deffayet C., Pujolas O., Sawicki I., Vikman A.: 2010, {\it J. Cosmol. Astropart. Phys.}, {\bf 10}, 026.\\
Easson D.A. \& Vikman A.: 2016, arXiv:1607.00996 [gr-qc].\\
Feng B., Wang X., Zhang X.: 2005, {\it Phys. Lett. B}, {\bf 607}, 35.\\
Komatsu E. et al.: 2009, {\it Astrophys. J. Suppl. Ser.}, {\bf 180}, 330.\\
Linder E.V.: 2003, {\it Phys. Rev. Lett.}, {\bf 90}, 091301.\\
Novosyadlyj B. \& Sergijenko O.: 2009, {\it Journal of Physical Studies}, {\bf 13}, 1902.\\
Pettorino V. \& Baccigalupi C.: 2008, {\it Phys. Rev. D}, {\bf 77}, 103003.\\
Sergijenko O. \& Novosyadlyj B.: 2008, {\it Kinematics and Physics of Celestial Bodies}, {\bf 24}, 259.\\
Vikman A.: 2005, {\it Phys. Rev. D}, {\bf 71}, 023515.\\
\vfill
\end{document}